# Google Scholar Metrics 2013: nada nuevo bajo el sol

Álvaro Cabezas-Clavijo y Emilio Delgado-López-Cózar

EC3 Research Group & EC3metrics Spin-Off, Universidad de Granada, Granada (Spain)

**RESUMEN**

Se presentan las características más significativas de la nueva versión de Google Scholar Metrics (julio 2013), subrayando las novedades y las debilidades detectadas en un primer análisis "de urgencia". Cabe destacar precisamente la ausencia de novedades respecto a anteriores actualizaciones, ya que la única modificación es la actualización del marco temporal (2008-2012); por tanto persisten los problemas enunciados en trabajos previos. Finalmente, parece confirmarse que Google actualizará este producto con carácter anual.

**PALABRAS CLAVE**

Google Scholar / Google Scholar Metrics / Revistas / Citas / Bibliometría / Índice H / Evaluación / Rankings

# Google Scholar Metrics 2013: nothing new under the sun

**ABSTRACT**

Main characteristics of Google Scholar Metrics' new version (july 2013) are presented. We outline the novelties and the weaknesses detected after a first analysis. As main conclusion, we remark the lack of new functionalities with respect to last editions, as the only modification is the update of the timeframe (2008-2012). Hence, problems pointed out in our last reviews still remain active. Finally, it seems Google Scholar Metrics will be updated in a yearly basis.

**KEYWORDS**

Google Scholar / Google Scholar Metrics / Journals / Citations / Bibliometrics / H index / Evaluation / Rankings

# INTRODUCCIÓN

Acaba de hacerse pública la versión 2013 de Google Scholar Metrics. Tenemos por fuerza que comenzar repitiendo lo que ya dijimos en nuestra revisión anterior (1) y es que en lo que empieza a ser ya la seña de identidad de esta empresa, Google ha decidido sorprendernos con el lanzamiento de la actualización de Google Scholar Metrics, en este caso, el 25 de julio, en plenas vacaciones veraniegas. Parece que Google se ha empeñado en seguir siendo diferente, y si las empresas dedicadas a la evaluación bibliométrica de revistas (Thomson Reuters y Scopus) actualizan sus productos anualmente, Google lo ha hecho esta vez 8 meses y medio después del lanzamiento de la segunda versión del producto (15 de noviembre 2012). Recordemos que la anterior vez fue a los 7 meses y medio de su nacimiento el 1 de abril de 2012. Sin embargo, y a tenor de lo expuesto en la nota de lanzamiento parece que ésta será la única versión que se lance en 2013, lo que sugiere la actualización de este producto cobrará carácter anual.

Hemos de confesar que esperábamos el movimiento de la compañía californiana. La actualización del producto de Google era un paso lógico tras el decepcionante producto inicial (2,3) y la posterior mejora sustancial del mismo (1). Esta mejora apuntada en noviembre de 2012 parece, sin embargo, que no se ha visto refrendada en esta actualización de julio de 2013 ya que no se aprecian novedades o mejoras respecto a la versión previa, excepción hecha de la supresión del ranking de revista en coreano. En definitiva,; **Google se ha limitado a actualizar los datos**. Por tanto, persisten muchos de los errores y limitaciones señalados en trabajos previos como la falta de categorización por áreas y disciplinas de las revistas no inglesas, los problemas de normalización o las dificultades para enlazar a la fuente primaria de un trabajo.

En esta nota *urgente* repasamos las características más significativas de esta nueva versión de Google Scholar Metrics, subrayando las novedades y las debilidades detectadas.

# DESCRIPCIÓN

Scholar Metrics sigue ofreciendo en esta versión la clasificación de las primeras 100 revistas del mundo por idioma de publicación, ordenadas según el índice h y la mediana del número de citas obtenida por los artículos que contribuyen al índice h. El período sobre el que se computa el indicador se mantiene en los últimos cinco años (2008-2012), solo que ahora el cálculo basado en las citas emitidas por los artículos indizados en Google Scholar se extiende hasta mitad de julio de 2013.

*¿Qué se mantiene?*

- Solo se incluyen revistas que hayan publicado al menos 100 artículos en un periodo de cinco años (2008-2012 en este caso) y que hayan recibido alguna cita (esto es, se excluyen las revistas con índice h=0).
- De cada revista solo se visualizan los artículos que contribuyen al índice h (pudiéndose consultar los documentos citantes (pinchando en *cited by*)
- Asimismo se pueden buscar revistas por palabras incluidas en el título usando el buscador. En este caso, solo se ofrecen 20 resultados.
- Los rankings de revistas se presentan por idiomas (en este caso nueve: inglés, chino, portugués, alemán, español, francés, japonés, holandés e italiano). Respecto a la versión anterior, se ha excluido el ranking de revistas en coreano. Por cada idioma se presenta un listado de las 100 revistas con mayor índice h, excepción hecha de las revistas inglesas donde se agrupan, además, por ocho áreas temáticas (*Humanities, Literature & Arts, Social Sciences,...*) y 313 disciplinas (Religion, *Language &*

Google Scholar Metrics 2013: nada nuevo bajo el sol.

*Linguistics, History, Algebra,...*[1]). Lamentablemente esta opción solo existe para las revistas en inglés y sólo se muestran las 20 revistas con mayor índice h en cada una de las categorizaciones realizadas.

Las revistas pueden estar clasificadas en varias áreas o disciplinas temáticas, si bien lo habitual es que cada revista se clasifique únicamente en un área y en una disciplina. Cuando se pincha en su índice h puede saberse en qué disciplinas figura (siempre y cuando esté entre las veinte primeras), y cuál es su posición.

En la página de ayuda, se indica con mayor nitidez cuales son las fuentes cubiertas por Google Scholar Metrics (artículos de revistas procedentes de las webs que cumplen los criterios de inclusión de Google, algunas series de congresos en Informática e Ingeniería Eléctrica, así como preprints alojados en algunos repositorios seleccionados) y las no cubiertas (documentos judiciales, tesis, libros y patentes). Mantener la indización de congresos es una medida sumamente acertada pues hay áreas (*Computational Linguistics, Computer Graphics, Computer Hardware Design, Computer Networks & WirelessCommunication, Computer Security & Cryptography, Computer Vision & Pattern Recognition, Computing Systems, Data Mining & Analysis,* o *Databases & InformationSystems*) (figura 1) donde estos juegan un papel crucial en la comunicación científica, aspecto al que dedicaremos próximamente un análisis detallado.

*Figura 1*
*Presencia de Congresos en los rankings de algunas especialidades de Computación*

| | Publication | h5-index | h5-median |
|---|---|---|---|
| 1. | Meeting of the Association for Computational Linguistics (ACL) | 61 | 77 |
| 2. | Conference on Empirical Methods in Natural Language Processing (EMNLP) | 49 | 76 |
| 3. | North American Chapter of the Association for Computational Linguistics | 39 | 51 |
| 4. | International Conference on Computational Linguistics (COLING) | 35 | 50 |
| 5. | Language Resources and Evaluation (LREC) | 35 | 50 |
| 6. | Computational Linguistics | 28 | 72 |
| 7. | Computer Speech & Language | 28 | 41 |
| 8. | Workshop on Statistical Machine Translation | 25 | 41 |
| 9. | Conference of the European Chapter of the Association for Computational Linguistics (EACL) | 24 | 35 |
| 10. | arXiv Computation and Language (cs.CL) | 21 | 45 |
| 11. | International Joint Conference on Natural Language Processing | 21 | 34 |

Para el caso de los repositorios se mantiene asimismo la opción de computar sólo aquellas colecciones de documentos dentro de un repositorio sobre los que se pueda calcular indicadores bibliométricos dejando de aparecer *Arxiv, RePec* y *SSRN* en su conjunto, lo cual generaba, además, una indeseable distorsión en el cálculo del índice h (4). Dichos repositorios juegan un importante papel en áreas como la Física (figura 2) o la Economía (figura 3)

---

[1] El anexo con las áreas y disciplinas establecidas puede encontrarse en http://acabezasclavijo.files.wordpress.com/2012/11/google-scholar-areas-disciplines.pdf

*Figura 2*
*Presencia de repositorios en los rankings de algunas especialidades de Física*

Top publications - High Energy & Nuclear Physics   Learn more

| Publication | h5-index | h5-median |
|---|---|---|
| 1. **arXiv High Energy Physics - Phenomenology (hep-ph)** | 141 | 185 |
| 2. Physical Review D | 133 | 185 |
| 3. arXiv High Energy Physics - Theory (hep-th) | 133 | 180 |
| 4. Journal of High Energy Physics | 125 | 174 |
| 5. arXiv High Energy Physics - Experiment (hep-ex) | 92 | 136 |
| 6. Physics Letters B | 92 | 124 |

Top publications - Physics & Mathematics   Learn more

| Publication | h5-index | h5-median |
|---|---|---|
| 1. Physical Review Letters | 197 | 262 |
| 2. The Astrophysical Journal | 149 | 198 |
| 3. **arXiv High Energy Physics - Phenomenology (hep-ph)** | 141 | 185 |
| 4. arXiv Superconductivity (cond-mat.supr-con) | 140 | 209 |
| 5. arXiv Materials Science (cond-mat.mtrl-sci) | 133 | 221 |
| 6. Physical Review D | 133 | 185 |
| 7. arXiv High Energy Physics - Theory (hep-th) | 133 | 180 |
| 8. Physical Review B | 132 | 178 |
| 9. arXiv Mesoscale and Nanoscale Physics (cond-mat.mes-hall) | 130 | 202 |
| 10. arXiv Cosmology and Extragalactic Astrophysics (astro-ph.CO) | 130 | 180 |
| 11. arXiv Quantum Physics (quant-ph) | 127 | 195 |

*Figura 3*
*Presencia de repositorios en los rankings de algunas especialidades de Economía*

Top 20 publications matching *working papers*   Learn more

| Publication | h5-index | h5-median |
|---|---|---|
| 1. NBER Working Papers | 161 | 222 |
| 2. Policy Research Working Paper Series | 64 | 97 |
| 3. European Central Bank Working Paper Series | 56 | 88 |
| 4. World Bank Policy Research Working Paper Series | 55 | 91 |
| 5. CESifo Working Paper Series | 53 | 86 |
| 6. Bank for International Settlements Working Papers | 43 | 88 |
| 7. Bank of Italy Temi di Discussione (Working Paper) | 39 | 69 |
| 8. HAL Working Papers | 38 | 65 |
| 9. Federal Reserve Bank of San Francisco Working Paper Series | 37 | 75 |
| 10. Harvard Business School Working Papers | 36 | 56 |
| 11. Department of Economics and Business, Universitat Pompeu Fabra Economics Working Papers | 36 | 53 |
| 12. Harvard University, John F. Kennedy School of Government Working Paper Series | 35 | 66 |

### *¿Cuáles son las novedades?*

**NINGUNA** en cuanto a los aspectos técnicos, de prestaciones o de procesamiento de la información; la única novedad y más importante es que Google nos ha sacado de dudas y parece que mantiene este servicio y la derivada de los posibles cambios en los diferentes rankings por idiomas, y entrada de nuevas revistas a las distintas clasificaciones. A simple vista, parecen pocas las nuevas revistas que aparecen en los rankings, confirmando la estabilidad que ya habíamos detectado al comparar las dos versiones publicadas en 2012 (5) Asimismo la nota de lanzamiento de esta versión[2] parece confirmar el carácter anual de la misma "[…] we are releasing the 2013 version of Scholar Metrics". Esto sugiere que podría mantenerse a partir de ahora este carácter anual en la actualización de los datos, algo habitual en los productos bibliométricos, si bien la clásica opacidad de Google nos impide ir más allá de las meras conjeturas.

## LIMITACIONES

Las principales limitaciones en esta nueva versión de Google Scholar Metrics son las siguientes:

- Desglose por áreas y disciplinas únicamente para publicaciones en inglés. Para el resto de idiomas se han actualizado los indicadores pero sigue siendo imposible acceder a las revistas más destacadas por categorías. Este hecho es incoherente con una de las principales virtudes tradicionales de Google Scholar como es el haber dado visibilidad las revistas no incluidas en la corriente principal de la ciencia.
- Falta de transparencia en la categorización temática realizada por la empresa, ya que sigue sin explicitar ni el criterio seguido para conformar las áreas y disciplinas ni el criterio para la clasificación de las revistas en un área o disciplina. Como dato positivo, se han subsanado errores de bulto señalados en nuestra anterior revisión, como la no clasificación de la revista *Journal of the American Society for Information Science and Technology* en la disciplina *Library &InformationScience.* Subsanado este error, JASIST lidera la clasificación de este área. También se ha modificado la inconsistente clasificación de las revistas de Economía, previamente incluidas tanto en la categoría "Ciencias Sociales" como en "Empresa, Economía y Administración". Ahora están registradas en la segunda de las categorías mencionadas.
- Imposibilidad de acceder a las versiones anteriores de Google Scholar Metrics, con lo que no se puede analizar la evolución de la revista por marcos temporales. Esta circunstancia, unida al desconocimiento sobre las fechas de actualización del producto y sobre los marcos temporales y ventanas de citación que Google establece para construir el mismo, puede provocar inconsistencia en los indicadores generados y confusión a la hora de interpretar los resultados obtenidos. A este respecto la propia empresa señala en su nota de lanzamiento que el hecho de que el marco temporal sea algo más corto en esta ocasión (4 meses menos) puede generar un pequeño descenso en los índices de impacto de las revistas.
- No se especifica en cuántas series o colecciones se han dividido los distintos repositorios.
- Visualización únicamente de las 20 primeras revistas de cada área y disciplina. No es posible conocer pues la posición el resto de publicaciones.
- Por último, en cuanto al control bibliográfico de la información en Scholar Metrics, no parece haberse avanzado mucho en estos últimos meses, ya que no se han eliminado totalmente los títulos duplicados. Es muy curioso que estos duplicados afecten en este caso a las dos publicaciones más relevantes de la Documentación española como son *El profesional de la Información* y la *Revista Española de Documentación Científica*, ya que ambas aparecen por duplicado (figura 4).

---

[2] 2013 Scholar Metrics released http://googlescholar.blogspot.com.es/2013/07/2013-scholar-metrics-released.html

**Figura 4**
*El Profesional de la Información, y Revista Española de Documentación Científica, por duplicado*

| Publicación | Índice h5 | Mediana h5 |
|---|---|---|
| 1. El Profesional de la Informacion | 16 | 21 |
| 2. PROFESIONAL DE LA INFORMACION | 6 | 8 |

Publicaciones que coinciden con *Profesional Informacion*

*Las fechas y los recuentos de citas son estimados y se determinan de forma automática mediante un programa informático.*

Publicaciones que coinciden con *española Documentación Cientifica*

| Publicación | Índice h5 | Mediana h5 |
|---|---|---|
| 1. Revista española de Documentación Científica | 10 | 12 |
| 2. Revista Espanola de Documentacion Cientifica | 5 | 11 |

*Las fechas y los recuentos de citas son estimados y se determinan de forma automática mediante un programa informático.*

Por otra parte, se mantienen errores en la búsqueda y localización de revistas, que si bien no son numerosos, sí comprometen el rigor del producto final (figura 5).

**Figura 5**
*Errores en la indización de revistas*

Publicaciones que coinciden con *63*

| Publicación | Índice h5 | Mediana h5 |
|---|---|---|
| 1. br/>< b> Warning</b>: get_class () expects parameter 1 to be object, array given in< b> D:\ wamp\ www\ www. eaafj. or. ke\ classes\ cache\ GenericCache. inc. php</b> on line< b> 63</b>< br/> East African Agricultural and Forestry Journal | 2 | 2 |

*Las fechas y los recuentos de citas son estimados y se determinan de forma automática mediante un programa informático.*

Publicaciones que coinciden con *vol. 1*

| Publicación | Índice h5 | Mediana h5 |
|---|---|---|
| 1. International Journal of Advancements in Research and Technology, vol. 1, no. 6, p. 101-106 | 1 | 2 |

Muy evidentes son los problemas que se detectan en el enlazado de documentos. Google ha demostrado ser capaz de agrupar con bastante acierto las distintas versiones de un documento (publicadas o almacenadas en revistas, repositorios o web académicos y de investigación de todo tipo), sin embargo, no consigue remitir a la versión original de dichos documentos, a pesar de que la empresa declare que esta es su política. Como botón de muestra sirvan los ejemplos de European Central Bank Working Paper Series que dirige a http://www.ssrn.com/ y no a su sede principal

Google Scholar Metrics 2013: nada nuevo bajo el sol.

http://www.ecb.europa.eu/pub/scientific/wps/date/html/index.en.html. Lo mismo ocurre con otras series de working papers (*World Bank Policy Research WorkingPaper Series, CESifo WorkingPaper Series, Bank for International Settlements Working Papers, Bank of Italy Temi di Discussione (WorkingPaper), Federal Reserve Bank of San Francisco WorkingPaper Series, Harvard Business School WorkingPapers, Department of Economics and Business, Universitat Pompeu Fabra Economics Working Papers*). Igual fenómeno sucede con muchas revistas, especialmente aquellas que no pertenecen a la corriente principal de la ciencia, que teniendo versión electrónica (véase Psicothema, por ejemplo en figura 6) remite a la página de fuentes secundarias, como Pubmed (figura 6), incumpliendo abiertamente lo afirmado por Google.

**Figura 6**
**Versión electrónica de un artículo en Psicothema y enlace a la referencia en Pubmed de dicho artículo que realiza Google Scholar Metrics**

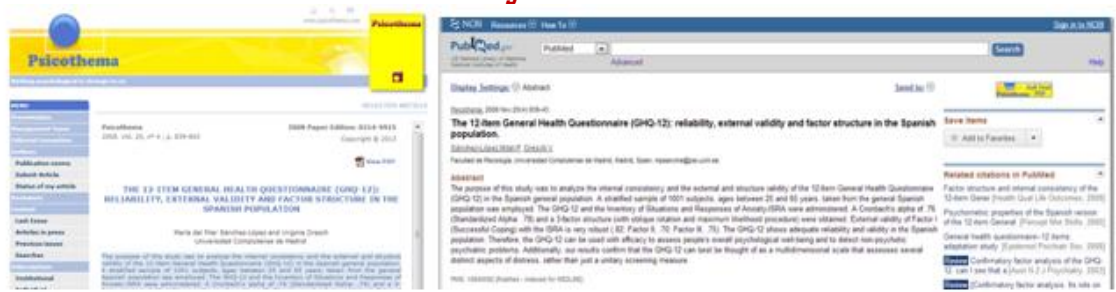

Las revistas poseen otro tipo de problemas. El primero de ellos es que muchos de sus artículos se quedan sin enlazar, a pesar de que la revista posee versión electrónica (figura 7).

**Figura 7**
**Artículos sin enlace a fuente de datos**

Archivos de bronconeumología

h5-index: **20**  h5-median: **24**

| Title / Author | Cited by | Year |
|---|---|---|
| Guía clínica SEPAR-ALAT de diagnóstico y tratamiento de la EPOC<br>G Peces-Barba, J Albert Barberà, À Agustí, C Casanova, A Casas, J Luis ...<br>Archivos de Bronconeumología 44 (5), 271-281 | 110 | 2008 |
| Diagnosis and Treatment of Bronchiectasis<br>M Vendrell, J de Gracia, C Olveira, MÁ Martínez, R Girón, L Máiz, R Cantón ...<br>Archivos de Bronconeumología ((English Edition)) 44 (11), 629-640 | 81 | 2008 |
| Diagnóstico y tratamiento de la tuberculosis<br>J Ruiz-Manzano, R Blanquer, J Luis Calpe, JA Caminero, J Caylà, JA Domínguez ...<br>Archivos de Bronconeumología 44 (10), 551-566 | 79 | 2008 |
| Documento de consenso sobre diagnóstico, tratamiento y prevención de la tuberculosis<br>J González-Martín, JM García-García, L Anibarro, R Vidal, J Esteban, R ...<br>Archivos de Bronconeumología 46 (5), 255-274 | 47 | 2010 |
| Recomendaciones para el tratamiento farmacológico del tabaquismo. Propuestas de financiación<br>CA Jiménez-Ruiz, JA Riesco Miranda, Á Ramos Pinedo, M Barrueco Ferrero, S ...<br>Archivos de Bronconeumología 44 (4), 213-219 | 39 | 2008 |
| Diagnosis and management of chronic obstructive pulmonary disease: joint guidelines of the Spanish Society of Pulmonology and Thoracic Surgery (SEPAR) and the Latin American Thoracic Society (ALAT)].<br>G Peces-Barba, JA Barberà, A Agustí, C Casanova, A Casas, JL Izquierdo, J ...<br>Archivos de bronconeumología 44 (5), 271 | 35 | 2008 |
| Estándares asistenciales en hipertensión pulmonar: Documento de consenso elaborado por la Sociedad Española de Neumología y Cirugía Torácica (SEPAR) y la Sociedad Española de Cardiología (SEC)<br>JA Barberà, P Escribano, P Morales, M Ángel Gómez, M Oribe, Á Martínez, A ...<br>Archivos de Bronconeumología 44 (2), 87-99 | 33 | 2008 |

Por otra parte, sigue siendo incapaz de fusionar las revistas que se editan en dos idiomas (figura 8)

*Figura 8*
*Revistas con dos versiones en distintos idiomas procesadas de forma diferenciada*

| Publicación | Índice h5 | Mediana h5 |
|---|---|---|
| 1. Anales de Pediatría | 14 | 17 |
| 2. Anales de pediatría (Barcelona, Spain) | 10 | 14 |
| 1. Farmacia hospitalaria | 10 | 11 |
| 2. Farmacia Hospitalaria (English Edition) | 6 | 7 |
| 1. Reumatología Clínica | 8 | 12 |
| 2. Reumatología Clínica (English Edition) | 6 | 6 |

Por último, otro problema no resuelto es el de las incongruencias que se producen en la búsqueda de revistas por palabras, unas incoherencias impropias de un buscador de la categoría de Google. Es un despropósito que una búsqueda con una determinada raíz no sea capaz de recuperar las revistas con mayor índice h en los 20 primeros resultados. Cuando se busca con las raíces "cardiol" o "cardiolog" se recuperan una lista de revistas (figura 9) donde no figuran la Revista Española de Cardiología o Arquivos Brasileiros de Cardiologia, que, con la misma raíz, poseen índices h de 28 y 24, superiores pues, a las mostradas en la búsqueda con dichas palabras.

*Figura 9*
*Búsquedas de revistas por raíz léxica con resultados inconsistentes*

cardiol

20 publicaciones principales que coinciden con *cardiol*   Más información

| Publicación | Índice h5 | Mediana h5 |
|---|---|---|
| 1. Journal of the American College of Cardiology | 160 | 211 |
| 2. The American Journal of Cardiology | 69 | 86 |
| 3. Journal of Molecular and Cellular Cardiology | 56 | 75 |
| 4. Nature Reviews Cardiology | 52 | 81 |
| 5. International Journal of Cardiology | 51 | 71 |
| 6. Basic Research in Cardiology | 42 | 63 |
| 7. Clinical Research in Cardiology | 34 | 46 |
| 8. Journal of Nuclear Cardiology | 34 | 42 |
| 9. Canadian Journal of Cardiology | 33 | 46 |
| 10. Current Opinion in Cardiology | 31 | 38 |
| 11. Pediatric Cardiology | 26 | 33 |
| 12. Clinical Cardiology | 26 | 32 |
| 13. Cardiology | 25 | 34 |
| 14. Cardiology in Review | 25 | 32 |
| 15. Journal of Cardiology | 21 | 29 |

# CONCLUSIONES

Google Scholar Metrics 2013: nada nuevo bajo el sol.

Como reza en el título de esta apresurada revisión del nuevo producto, parece que Google ha optado por dar estabilidad a su producto de evaluación de revistas científicas a través de los recuentos de citas, sin añadir ninguna mejora aparente. Es loable la corrección de algunos errores de bulto denunciados en anteriores trabajos, y continúa siendo apreciable el concurso humano (no sólo de robots) en la definición de áreas, disciplinas y en la categorización de revistas. Sin embargo, Scholar Metrics sigue distando de ser una herramienta fiable y válida debido a las múltiples limitaciones de índole metodológico señaladas. Pero especialmente, algo que hemos venido denunciando insistentemente: es sensible a las manipulaciones y no es lo suficientemente transparente para que la comunidad científica pueda detectarlas (6,7). En este sentido, resulta imprescindible que se ofrezca el índice H con autocitas y sin autocitas, que suele ser la triquiñuela más a mano de autores o editores para inflar el impacto de la revista.

Éstas son unas primeras valoraciones de urgencia de la actualización 2013 de Google Scholar Metrics. En próximos días ofreceremos un análisis más detallado del producto y de las consecuencias que se derivan para la evaluación investigadora.

## AGRADECIMIENTOS

## MATERIAL SUPLEMENTARIO

Índice de áreas y disciplinas incluidas en Google Scholar Metrics
http://acabezasclavijo.files.wordpress.com/2012/11/google-scholar-areas-disciplines.pdf

## REFERENCIAS BIBLIOGRÁFICAS

## OTROS TRABAJOS DE EC3 SOBRE GOOGLE SCHOLAR